\documentclass[10pt,letterpaper,twocolumn]{article}

\usepackage{ol2}
\usepackage{amsmath,times,epsfig,graphicx}

\begin{document}

\twocolumn[

\title{Miniature Optical Atomic Clock: Stabilization of a Kerr Comb Oscillator}

\author{A. A. Savchenkov, D. Eliyahu, W. Liang, V. S. Ilchenko, J. Byrd, A. B. Matsko$^*$,  D.  Seidel,  L. Maleki }

\address{OEwaves Inc., 465 North Halstead Street, Suite 140, Pasadena, CA 91107 \\
$^*$Corresponding author: andrey.matsko@oewaves.com }

\begin{abstract}
Mechanical clocks consist of a pendulum and a clockwork that translates the pendulum period to displayed time. The most advanced clocks utilize optical transitions in atoms in place of the pendulum and an optical frequency comb generated by a femtosecond laser as the clockwork. The comb must be stabilized at two points along its frequency spectrum: one with a laser to lock a comb line to a transition in the atom, and another through self referencing to stabilize the frequency interval between the comb lines. This approach requires advanced techniques, so optical atomic clocks are currently laboratory devices in specialized labs. In this paper we leverage unique properties of Kerr comb oscillators for realization of optical atomic clocks in miniature form factors. In particular, we describe a clock based on D$_1$ transition of $^{87}$Rb that fits in the palm of the hand, and can be further miniaturized to chip scale.
\end{abstract}


 ]

A two-point frequency stabilization for a femtosecond laser-based frequency comb is fundamentally necessary, since the phase of the carrier envelope and pulse repetition rate of a mode locked laser are independent variables \cite{hall04ch}. An octave-wide frequency comb is often required to lock a line of the comb to its second harmonic to stabilize the carrier envelop phase.  This is in addition to a separate lock of the comb to a laser that is, in turn, locked to an atomic transition. The laser must also be locked to a high finesse optical cavity to reduce its linewidth.  In this Letter we demonstrate a new and far simpler clock architecture utilizing a microresonator-based Kerr frequency comb.

The optical Kerr comb is the byproduct of $\chi^{(3)}$ nonlinearity of the material of which the resonator is made, and is generated by pumping one of the resonator modes with a continuous wave (cw) laser \cite{delhaye07n,savchenkov08prl}. Kerr comb generation is fundamentally a unitary process, and can occur in a resonator where the only losses are related to out  coupling of the pump light. In such a case the repetition frequency of the comb does not depend on either frequency or power of the pump light, and the comb becomes an ideal oscillator that does not require any stabilization.  The stability of the repetition rate of such a  comb is given by a quantum Shawlow-Townes-like formula \cite{matsko05pra}.  In practice, contact of the resonator with the environment and finite absorption in the resonator host material spoil the stability conditions, and the repetition rate of the comb becomes dependent on parameters of the pump light. Locking the pump laser to an atomic cell and locking the resonator to the laser becomes necessary to regain the stability.

There have been previous successful attempts for frequency stabilization of Kerr frequency comb oscillators. In two experiments, the comb was locked to a reference femtosecond optical frequency comb \cite{delhaye08prl,delhaye12arxiv}. Locking the repetition rate of a Kerr comb to a reference RF signal has also been demonstrated \cite{papp12arxiv}. In all these studies, the absolute frequency stability of the comb repetition rate resulted in  Allan deviation of $10^{-12}$ at 1~s, with subsequent longer term stability exhibiting $\tau^{-1}$ behavior, where $\tau^{-1}$ is the integration time. In this Letter we describe an optical clock based on the Kerr optical comb locked to optical transition in $^{87}$Rb with a semiconductor laser that is injection locked to the resonator. The physics package of this clock fits in the palm of a hand, and can be further miniaturized, down to chip scale. In contrast with previous work for Kerr frequency comb stabilization, our device does not require an additional master oscillator to gain its stability.

%
 \begin{figure}
 \center{
\epsfig{file=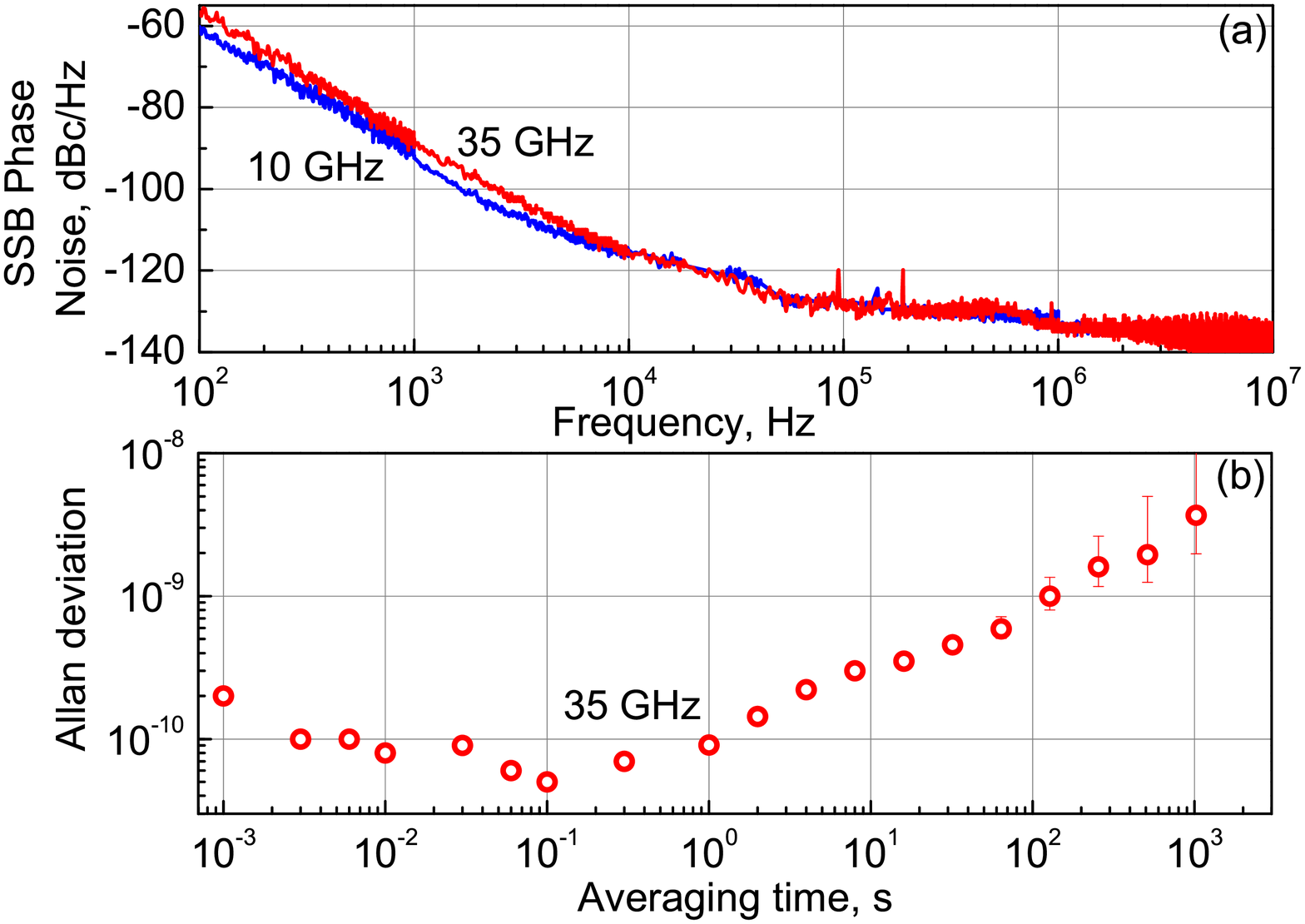, width=6.5cm, angle=0} }
\caption{\label{fig1} {\small Characteristics of a free running Kerr comb. Single sideband (SSB) phase noise, (a), of the 10 GHz as well as 35~GHz RF signal generated by free running optical Kerr frequency combs on a fast photodiode. The Allan deviation for the 35~GHz comb is shown in panel (b). }
}
 \end{figure}
%

A key feature of the Kerr comb is that it is enough to stabilize only a single point in its frequency spectrum to obtain stable clock operation \cite{maleki10ifcs,maleki11ifcs}. In this system the frequency of one comb line coincides with the frequency of the pump laser, while the comb repetition rate is fixed by the properties of the monolithic resonator generating the comb. Thus, the entire Kerr comb is stabilized once the frequency of the pump laser is stabilized to an external reference. Remarkably, there is no need to generate an octave spanning Kerr comb to stabilize it. In this respect, it is in principle enough to have only two harmonics in the comb (a.k.a. a hyper-parametric oscillator \cite{kippenberg04prl,savchenkov04prl}) to realize a stabilized clock. It is even advantageous to a the narrower comb for the stabilization purpose since the narrow comb is not impacted by the variations of the group velocity dispersion as well as quality factors of the resonator modes that may cause undesirable correlation between the comb repetition rate and the pump laser noise.

We studied the behavior of a free running Kerr frequency comb oscillator to illustrate the correlations between frequency of the optical pump locked to a mode of a nonlinear resonator and comb repetition rate. This correlation enables the single frequency point locking of the comb. We fabricated a WGM resonator with 35~GHz free spectral range (FSR) from a commercially available z-cut MgF$_2$ optical window and pumped the resonator with light from a 1550~nm distributed feedback (DFB) laser \cite{liang11ol}. The Kerr comb generation was observed when the pump power exceeded 0.1~mW. We pumped the resonator with 5~mW of light and collected 2.8~mW of light exiting the resonator on a fast photodiode characterized with responsivity of 0.5~A$/$W. The demodulated comb generated a 35~GHz radio frequency (RF) signal with -22~dBm power at the output of the photodiode.

%
 \begin{figure}
 \center{
\epsfig{file=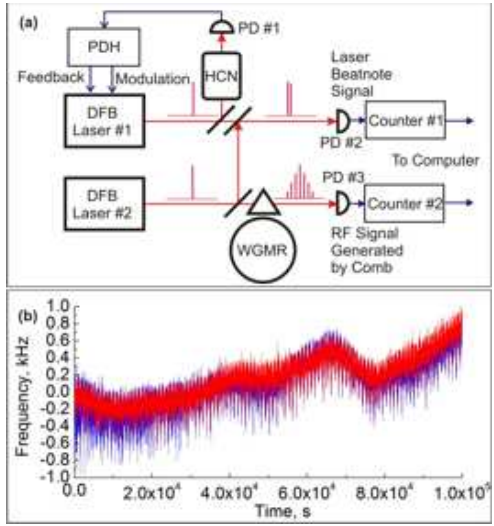, width=6.5cm, angle=0} }
\caption{\label{fig2} {\small Observation of correlation between drifts of the comb repetition rate and the pump frequency. (a) Schematic of the experimental setup used to observe correlation of the drifts of the optical comb harmonics as well as Kerr comb repetition rate. The comb repetition rate is measured directly via the RF signal generated at a fast photodiode. A separate laser locked to an HCN transition is utilized to measure the optical frequency. The beat frequency between two lasers is set to be approximately 50~MHz. (b) Simultaneously measured drifts of the optical frequency (blue) and RF frequency (red). The optical frequency drift is scaled by constant $\omega_0/\omega_{RF}$. }
}
 \end{figure}
%

It is especially challenging to obtain stable Kerr comb oscillation at lower RF frequencies since it requires usage of a larger monolithic resonator and its consequent larger mode density. This problem can be solved if ultra-high Q crystalline resonators are used since narrow bandwidth simplifies selective coupling to a particular mode. We fabricated a CaF$_2$ WGM resonator with 10~GHz FSR and $3\times10^9$ quality factor, pumped it with 5~mW of 1550~nm light, and demonstrated generation of a coherent Kerr comb.

We used an Agilent signal analyzer and a 40~GHz frequency counter, both locked to a reference Rb clock, to measure the Allan deviation of the K$_a$-band signal. In addition, we used an OEwaves phase noise measurement system to measure phase noise of the RF signal generated by the comb oscillator. The resultant phase noise and Allan deviation are shown in Fig.~\ref{fig1}a and \ref{fig1}b, respectively. The phase noise of the 10~GHz comb oscillator was approximately the same as the performance of the 35~GHz oscillator. The free running comb oscillator has an outstanding spectral purity as well as frequency stability as compared with any compact K$_a$ band oscillator. It is also worth noting that the data shown in Fig.~\ref{fig1} display nearly two orders of magnitude better performance than previously reported data obtained with Kerr comb oscillators.

%
 \begin{figure}
 \center{
\epsfig{file=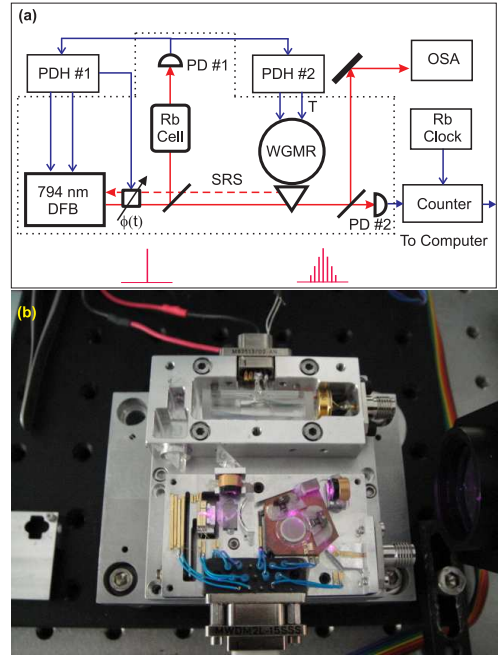, width=6.5cm, angle=0} }
\caption{\label{fig3} {\small Rubidium clock based on Kerr comb oscillator. (a) Schematic of the optical Rb clock based on the Kerr frequency comb oscillator. (b) Picture of the assembled physics package of the clock. The package includes the parts confined within the dashed line box shown in Fig.(a). }
}
 \end{figure}
%

The long term stability of the Kerr comb oscillator is limited because of the frequency drift resulting from  interaction  with the environment. One of the channels of the interaction is purely thermal. Even though the entire platform the oscillator is thermally stabilized with a thermoelectric element, the temperature of the resonator fluctuates within one degree throughout the day. Another channel is related to the dependence of resonator temperature on the laser pump power resulting from the residual optical absorption.

We used self-injection locking \cite{liang10ol} to achieve efficient pumping of the Kerr comb oscillator \cite{maleki10ifcs1} with a semiconductor laser. Self-injection locking results in collapse of the laser linewidth and also keeps the laser frequency tightly within the WGM passband, but still allows relative frequency walk between the laser and the pumped mode. The magnitude of this walk is comparable with the mode bandwidth, and the resultant changes in the optical detuning results in variation of the power circulating in the resonator. The power change modifies the absorbed power to change the temperature within the mode volume. We found that 0.1~mA change of the laser current in our prototype resulted in 1~kHz shift of the oscillation frequency.

%
 \begin{figure}
 \center{
\epsfig{file=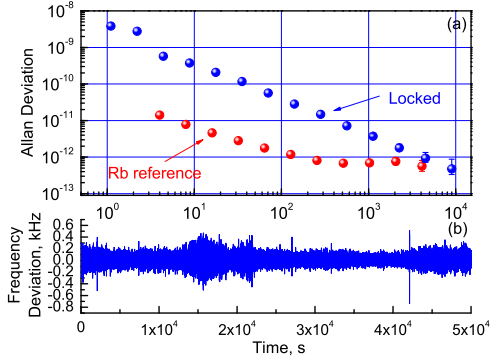, width=6.5cm, angle=0} }
\caption{\label{fig4} { \small  Performance of the Kerr frequency comb clock.  (a) Allan deviation plot for the clock (blue circles) and the smallest Allan deviation that can be measured with the setup (red circles). (b) Frequency measurement results used to calculate the Allan deviation of the clock. }
}
 \end{figure}
%

To establish that a stabilized clock could indeed be realized with this scheme, we must verify that the change in the RF frequency, $\omega_{RF}$, corresponds to the change of the optical frequency, $\omega_0$. The single point stabilization of the Kerr comb would only be possible if the optical drift is related to the RF drift by their frequency ratio  $\omega_0/\omega_{RF}$ at any moment of time. To examine this, we simultaneously measured both the optical frequency of the pump light  and the RF frequency generated by the comb.  We locked a 1550 nm DFB laser to the P9 molecular line of HCN in a sealed cell and used the laser as a reference oscillator (Fig.~\ref{fig2}a) to measure the frequency of the pump laser involved in frequency comb generation. The laser locked to the cell had stability (Allan deviation) of $3\times10^{-11}$  at 50~s averaging time ($\tau$), beyond which the Allan deviation started to drift as $\tau^{1/2}$. This reference stability was adequate to observe any drift of the  laser that was locked to the resonator mode.

We then changed the current of the pump laser in a controlled way and observed simultaneous shifts of the pump laser frequency and comb repetition rate (RF frequency), with the magnitudes of the shifts following the frequency ratio $\omega_0/\omega_{RF}$.  The RF and optical frequencies were simultaneously recorded for one day and then compared. The result of the measurement is shown in (Fig.~\ref{fig2}b). Again, the optical and RF frequencies drifted together and the ratio of their drifts was $\omega_0/\omega_{RF}$. This experiment confirmed that by reducing the drift of the optical frequency and locking it to an atomic transition,  the RF frequency produced by the comb could be stabilized.

The final step in demonstration of a Kerr comb based all-optical clock was locking of the pump laser to an atomic transition. We used saturated absorption dip in the D$_1$ line of $^{87}$Rb for this purpose; this feature is far narrower than the P9 molecular transition in HCN. The Kerr comb was generated in a high-Q CaF$_2$  WGM resonator pumped with a 794~nm DFB laser. The laser was then locked to the resonator mode via self-injection \cite{liang10ol} as well as a Pound-Drever-Hall (PDH) locking methods\cite{drever83apb}, and was simultaneously locked to the atomic transition.  This version of the all-optical clock had a physics package with dimensions $5\times6\times1.2$~cm (Fig.~\ref{fig3}b).

The clock's long term absolute stability was measured at about $3\times10^{-13}$ at $\sim 2\times10^4$~s integration time (Fig.~\ref{fig4}). The measured value was limited by the noise floor of the measurement setup determined by the stability of the commercial reference Rb clock.

Allan deviation in (Fig.~\ref{fig4}) averages as $\tau^{-1}$ since the clock stability is not limited by pure phase diffusion at small integration time. The resonator producing the comb is thermally modulated at $0.24$~Hz for locking the pump laser to the atomic transition. The modulation of the resonator temperature changes the comb repetition rate, so the clock signal is also modulated with the same frequency. The Allan deviation increases by two orders of magnitude at the averaging time corresponding to the modulation frequency, as compared with the free running comb oscillator. Averaging of the modulation signal results in decrease of the Allan deviation as $\tau^{-1}$. We expect that the averaging slows down when the phase diffusion limit is reached. The stability performance and Allan deviation of the clock was reproduced several times.

A miniature clock with the performance demonstrated here can find multiple applications in communications, sensing, and metrology.  The simplified architecture of the clock can also be used to achieve better stability using narrower transitions attainable with laser cooled and trapped atoms and ions.  Combined with these system, the simplified Kerr comb architecture can result in small all-optical clocks with extremely high stability and expand the use of these tools of precision metrology in many areas of science and technology.

The reported work was partially supported by DARPA. Authors acknowledge advice and guidance of Nathan P. Wells in the experimental techniques and measurements of the oscillator long term stability.

\end{document}